\documentclass[%
 reprint,
 showpacs,
 amsmath,amssymb,
 aps,
prstab,
]{revtex4-2}

\usepackage[english]{babel}
\usepackage{graphicx}
\usepackage{dcolumn}
\usepackage{bm}
\usepackage{siunitx}

\usepackage{array}
\newcolumntype{y}[1]{>{\let\newline\\\arraybackslash\hspace{0pt}}p{#1}}
\renewcommand*{\arraystretch}{1.2}

\begin{document}


\title{The Role of Low Intrinsic Emittance in Modern Photoinjector Brightness}

\author{Christopher M. Pierce}
\email{cmp285@cornell.edu}
\author{Matthew B. Andorf}
\author{Edmond Lu}
\affiliation{Cornell Laboratory for Accelerator-based Sciences and Education, Cornell University, Ithaca, New York 14853, USA}

\author{Matthew Gordon}
\author{Young-Kee Kim}
\affiliation{University of Chicago, Chicago, IL 60637, USA}

\author{Colwyn Gulliford}
\author{Ivan V. Bazarov}
\author{Jared M. Maxson}
\affiliation{Cornell Laboratory for Accelerator-based Sciences and Education, Cornell University, Ithaca, New York 14853, USA}

\author{Nora P. Norvell}
\author{Bruce M. Dunham}
\author{Tor O. Raubenheimer}
\affiliation{SLAC  National  Accelerator  Laboratory,  Menlo  Park,  CA,  USA}

\date{\today}

\begin{abstract}
Reducing the intrinsic emittance of photocathodes is one of the most promising routes to improving the brightness of electron sources.
However, when emittance growth occurs during beam transport (for example, due to space-charge), it is possible that this emittance growth overwhelms the contribution of the photocathode and, thus, in this case source emittance improvements are not beneficial.
Using multi-objective genetic optimization, we investigate the role intrinsic emittance plays in determining the final emittance of several space-charge dominated photoinjectors, including those for high repetition rate free electron lasers and ultrafast electron diffraction.
We introduce a new metric to predict the scale of photocathode emittance improvements that remain beneficial and explain how additional tuning is required to take full advantage of new photocathode technologies.
Additionally, we determine the scale of emittance growth due to point-to-point Coulomb interactions with a fast tree-based space-charge solver.
Our results show that in the realistic high brightness photoinjector applications under study, the reduction of thermal emittance to values as low as 50 pm/\si{\micro\meter} (1 meV MTE) remains a viable option for the improvement of beam brightness.
\end{abstract}

\maketitle

\section{Introduction}
Improving the brightness of space-charge dominated electron sources will unlock a wealth of next generation accelerator physics applications.
For example, the largest unit cell that may be studied with single shot ultrafast electron diffraction (UED) is limited by the beam's transverse coherence length, which is determined by transverse emittance, at a high enough bunch charge to mitigate the effects of shot noise in data collection.
The study of protein dynamics with UED requires producing $>1$ nm scale coherence lengths at more than $10^5$ electrons and sub-picosecond pulse lengths at the sample location \cite{siwick_femtosecond_2004, dwyer_femtosecond_2006}.
In another example, the intensity of coherent radiation available to the users of free electron lasers (FELs) is, in part, limited by beam brightness.
Beam brightness affects the efficiency, radiated power, gain length, and photon energy reach of FELs \cite{hyder_emittance_1988, di_mitri_estimate_2014}.

Photoinjectors equipped with low intrinsic emittance photocathodes are among the brightest electron sources in use today.
Peak brightness at the source is limited by two factors: the electric field at the cathode and the photocathode's transverse momentum spread.
Several short-pulse Child-Langmuir-like charge density limits have been derived for the photoemission regimes of relevance to practical photoinjectors.
These current density extraction limits make explicit the dependence of peak brightness on photocathode parameters and the electric field. \cite{bazarov_maximum_2009, filippetto_maximum_2014, shamuilov_child-langmuir_2018}.
Depending on the aspect ratio of the bunch, the brightness limit is super-linear in the electric field and motivates the push towards high accelerating gradient photoinjectors.
Contemporary DC, normal-conducting RF (NCRF), and superconducting RF (SRF) photoelectron guns have peak accelerating fields of order 10 MV/m \cite{pinayev_high-gradient_2015, dunham_performance_2007, dowell_status_2006, arnold_overview_2011} with very high repetition rates (well above 1 MHz).
At the cost of duty factor, state of the art NCRF electron guns can offer even higher fields of order 100 MV/m \cite{ferrario_homdyn_2000} and recent experimental results suggest the possibility of pushing peak fields to nearly 500 MV/m for cryogenically cooled accelerating structures \cite{rosenzweig_ultra-high_2018, cahill_high_2018, wang_experimental_1995, rosenzweig_next_2019, mceuen_high-power_1985, schwettman_low_1967, nordlund_defect_2012, fortgang_cryogenic_1987, descoeudres_dc_2009, grudiev_new_2009, dolgashev_geometric_2010, marsh_x_2011}.

In this work, we characterize the intrinsic emittance at the photocathode source via the Mean Transverse Energy (MTE):
\begin{equation}
    \varepsilon_{\text{C}} = \sigma_x \sqrt{\frac{\text{MTE}}{m c^2}},
    \label{eq:initial_emittance}
\end{equation}
where $\sigma_x$ is the laser spot size, and $m$ is the mass of the electron.
Here, it is clear that MTE plays the role of an effective temperature of emission.

\begin{figure*}[htp]
    \centering
    \includegraphics[width=\linewidth]{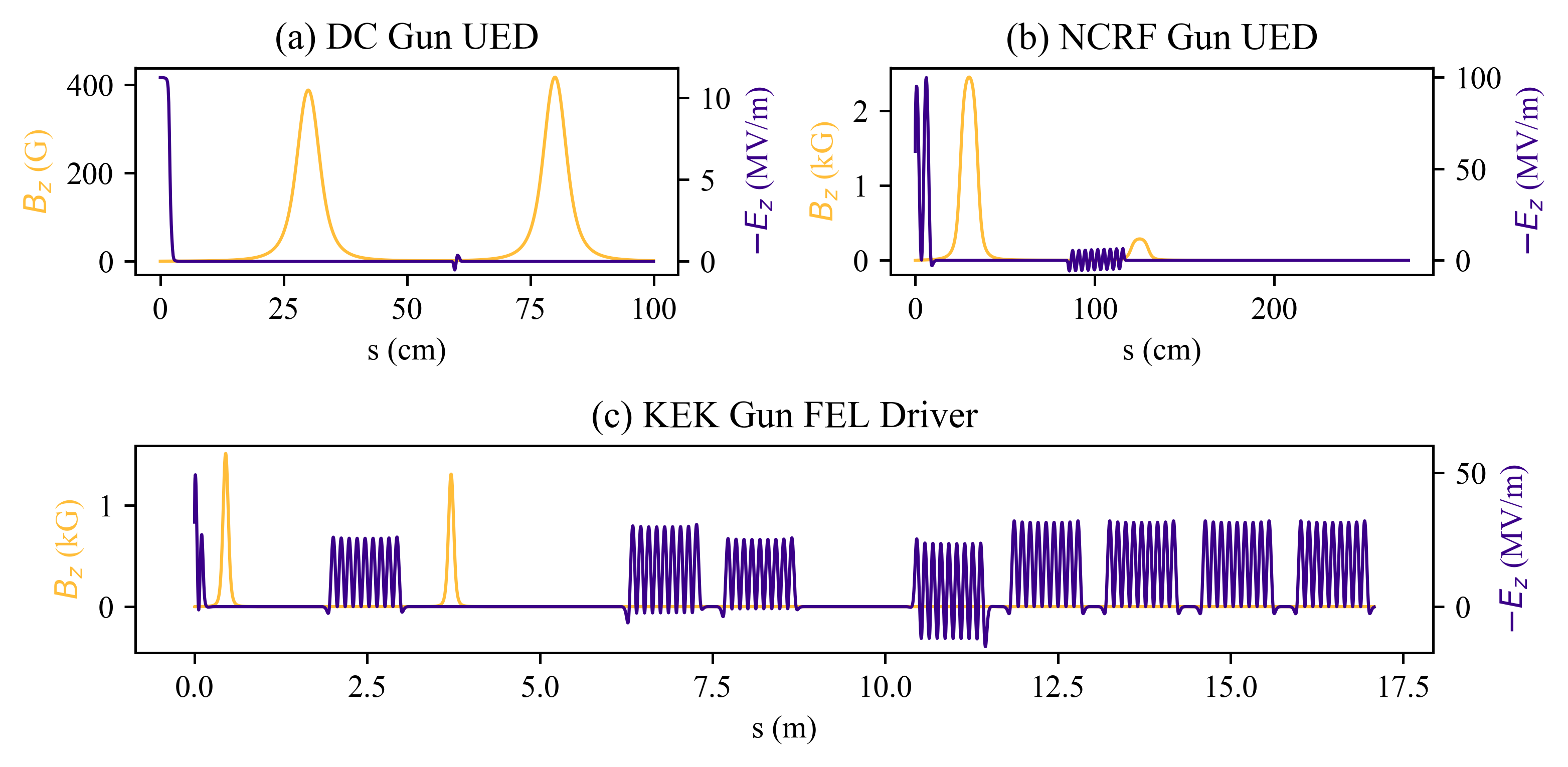}
    
    \caption{The on-axis electric and magnetic field as seen by a reference particle in the center of the electron bunch.  In each sub-figure, the cavity and magnet parameters are taken from an individual along the 0 meV Pareto front of the respective beamline.  Fields are output directly from General Particle Tracer and computed from ASTRA.}
    \label{fig:fields}
\end{figure*}

Great progress is being made in the discovery of low MTE photocathodes which are expected to improve the usable brightness of photoinjectors.
Due to the practical tradeoffs involved with photocathode choice, most photoinjectors today use materials with an MTE of around 150 meV \cite{weathersby_mega-electron-volt_2015, yang_100-femtosecond_2009, ding_measurements_2009, maxson_direct_2017}.
At the cost of QE, this MTE may be reduced by tuning the driving laser's wavelength.
For example, in  Cs$_3$Sb and Cs:GaAs photocathodes, the lowest MTE that may achieved via wavelength tuning at room temperature is nearly 35 meV and 25 meV respectively, but at $10^{-6}$ - $10^{-5}$ QE \cite{pastuszka_transverse_1997, cultrera_cold_2015, musumeci_advances_2018}.
Recent work has shown that the cryogenic cooling of photocathodes emitting at threshold can reduce MTE even further, potentially down to single digit meV MTEs \cite{karkare_<10_2018}.
However, a natural question arises amidst this progress in MTE reduction: in modern space-charge-dominated applications, to what extent does MTE reduction actually improve the final emittance?

Even in the case of linear transport, 3D space-charge effects lead to a transverse position-angle correlation which varies along the longitudinal coordinate and leads to an inflation of projected emittance that requires compensation \cite{carlsten_new_1988, floettmann_emittance_2017, qiu_demonstration_1996, serafini_envelope_1997}.
The residual emittance after compensation is due to non-linear forces, either from space-charge or beamline elements.
Scaling laws exist to help estimate their effects \cite{carlsten_space-charge-induced_1995, bazarov_comparison_2011}.
In some cases, non-linearity can cause phase space wave-breaking in unevenly distributed beams that is a source of irreversible emittance growth \cite{anderson_internal_1987, anderson_nonequilibrium_2000}.
Another irreversible cause of emittance growth is disorder induced heating (DIH) and other Coulomb scattering effects which are expected to become important in the cold dense beams of future accelerator applications \cite{maxson_fundamental_2013}.
Avoiding these emittance growth mechanisms requires the advanced design and tuning of photoinjector systems.

Multi-objective genetic algorithm (MOGA) optimization is a popular technique for the design and tuning of realistic photoinjectors \cite{baptiste_status_2009, panofski_multi-objective_2017, ineichen_massively_2013, panofski_multi-objective_2018, qian_s-band_2016, emery_global_2005, papadopoulos_multiobjective_2010}.
Photoinjectors often have to balance several key design parameters or objectives that determine the usefulness of the system for a given application.
MOGA is a derivative free method for computing the Pareto front, or family of highest performing solutions, in a parallel and sample efficient manner \cite{deb_fast_2002}.
Elitist genetic algorithms are known to converge to the global optima of sufficiently well-behaved fitness functions given enough evaluations \cite{rudolph_convergence_1996}.
This makes them well suited for problems involving many local extrema. 
Practical problems often require optimizations to be performed over a constrained search space and there exist techniques of incorporating these constraints into existing genetic algorithms without sacrificing efficiency \cite{bazarov_multivariate_2005}.

In this work, we examine the limits beam transport places on the ability of photoinjectors to take advantage of low MTE photocathodes in a diverse set of realistic simulated photoinjectors that have been tuned by a MOGA for ultimate performance.
This article begins with a discussion of our results involving the simulations of beamlines with idealized zero emittance photocathodes.
These simulations are performed on three important examples of high brightness electron beam applications: high repetition rate FELs, as well as single-shot DC and RF-based UED devices.
Using zero cathode emittance simulations, we introduce a new metric called the characteristic MTE to help understand the scale of photocathode MTE which is relevant to final beam quality.
It is shown that, depending on the properties of the beamline, system parameters need to be re-optimized to take full advantage of photocathode improvements.
We present a method of estimating when re-optimization needs to be performed and the magnitude of its effect on final emittance.
Finally, we set the scale for the magnitude of emittance growth due to point-to-point Coulomb interactions using a stochastic space-charge algorithm.

\section{Optimizations with a 0 \MakeLowercase{me}V MTE Photocathode}
To understand the contribution of photocathode MTE towards the final emittance of high brightness photoinjectors, we directly compare injector performance with a contemporary $\sim$150 meV MTE photocathode to what would be achievable with a perfect 0 meV MTE counterpart.
To cover the wide range of existing and near future accelerator technologies, we chose three realistic beamlines with significantly different energies as a representative set of high brightness photoinjector applications.
A DC and NCRF electron gun based single shot UED beamline reflect the two predominant energy scales of electron diffraction with single nanometer scale emittance at 10 - 100 fC bunch charge: order of magnitude 100 keV and 1 MeV.
At higher bunch charge, we select an SRF photoinjector under development at KEK expected to be capable of sub-$\mu$m scale emittance at 100 pC bunch charge for simulations representative of FEL driver applications.

The ultimate performance of each system is evaluated on the basis of the particle tracking codes General Particle Tracer \cite{van_der_geer_applications_1997} and ASTRA \cite{floettmann_astra:_2017} with optimization carried out in the framework of MOGA.
Children were generated with simulated binary crossover and polynomial mutation \cite{deb_multi-objective_2001}.
Selection was performed with SPEA-II \cite{zitzler_spea2:_2001} in the case of both UED examples and with NSGA-II \cite{deb_fast_2002} in the case of the FEL example.
Emittance preservation is known to depend strongly on the initial transverse and longitudinal distribution of the beam.
To this end, the optimizer is given the power to change parameters controlling the initial particle distribution using the same method described in \cite{bazarov_comparison_2011}.

The DC UED beamline is modeled after a similar system under development at Cornell University using the cryogenically cooled photoemission source described in \cite{lee_cryogenically_2018}.
The performance of this system under different conditions than presently considered is discussed in \cite{gulliford_multiobjective_2016}
where a detailed description of the layout and simulation methodology is also provided.
On-axis fields for this beamline are shown in Fig. \ref{fig:fields}a.
The beamline consists of two solenoids that surround an NCRF single cell bunching cavity and aid in transporting the high brightness beam to the sample located at $s = 1$ m.
The optimizer is given control over all magnet and cavity settings to minimize the RMS emittance at the sample while maximizing bunch charge.
Only solutions that keep the final spot size smaller than 100 \si{\micro\meter} RMS and the final beam length less than 1 ps RMS are considered.
These constraints were chosen based on common sample sizes used in diffraction \cite{weathersby_mega-electron-volt_2015} and the timescale of lattice vibration dynamics \cite{ligges_observation_2009, stern_mapping_2018}.
For a complete description of the decisions, objectives, and constraints used for this system, refer to Tab. \ref{tab:dc}.

The high gradient NCRF UED beamline is driven by a 1.6 cell 2.856 GHz gun capable of 100 MV/m and based on a design currently in use at a number of labs \cite{weathersby_mega-electron-volt_2015, musumeci_time_2009, zhu_femtosecond_2015, zhu_dynamic_2013, filippetto_design_2016}.
Samples are located at $s = 2.75$ m and the optimizer is given full control over two solenoids which surround a nine cell bunching cavity that is modeled after the first cell of the SLAC linac described in \cite{neal_stanford_1968}.
A discussion of our previous optimization experience with this beamline under a different set of constraints can be found in \cite{gulliford_multiobjective_2017}.
As in the case of the DC UED beamline, the optimizer was configured to minimize final RMS emittance while maximizing delivered bunch charge under the constraint of keeping the final spot size less than 100 \si{\micro\meter} RMS and the final length shorter than 1 ps RMS.
The decisions, objectives, and constraints of this optimization are detailed in Tab. \ref{tab:ncrf} and an example of the on-axis fields from an optimized individual is shown in Fig. \ref{fig:fields}b.

Our FEL driver example includes a 1.5 cell 1.3 GHz SRF gun in development at KEK for use in a CW ERL light source coupled with a photoinjector lattice aimed at use in the LCLS-II HE upgrade \cite{konomi_development_2019}.
The gun energy is controlled by the optimizer, but is in the range 1.5 - 3.5 MeV.   
Immediately after the gun is a 1.3 GHz 9 cell capture cavity surrounded by two solenoids.
The remaining cavities, of the same design as the capture cavity, are shown in the plot of external fields in Fig. \ref{fig:fields}c and accelerate the beam to its final energy of roughly 100 MeV.
Accelerating cavity number three was kept off during optimization as a planned backup for cavity failure in the real machine.
The bunch charge was fixed to 100 pC, and optimizations were performed to minimize both RMS emittance and bunch length at the end of the injector system. 
Energy constraints were tailored for the injector's use in the LCLS-II HE upgrade, and so we required valid solutions to have an energy greater than 90 MeV, an energy spread below 200 keV, and a higher order energy spread less than 5 keV.
The full set of decisions, objectives, and constraints is compiled in Tab. \ref{tab:fel}.

\begin{table}
\begin{tabular}{|y{145pt}|y{75pt}|}
\hline
{\bf Decision} & {\bf Range} \\
\hline\hline
Bunch Charge & 0 - 160 fC\\
Initial RMS Beam Size & 0 - 1 mm\\
Intitial RMS Beam Length & 0 - 50 ps\\
MTE & 0, 150 meV\\
Gun Voltage & 225 kV\\
Solenoid Current 1 and 2 & 0 - 4 A\\
Buncher Voltage & 0 - 60 kV\\
Buncher Phase & 90 degrees\\
\hline
\end{tabular}

\vspace{10pt}

\begin{tabular}{|y{145pt}|y{75pt}|}
\hline
{\bf Objective} & {\bf Goal} \\
\hline\hline
RMS Emittance & Minimize\\
Delivered Bunch Charge & Maximize\\
\hline
\end{tabular}

\vspace{10pt}

\begin{tabular}{|y{145pt}|y{75pt}|}
\hline
{\bf Constraint} & {\bf Value} \\
\hline\hline
Final RMS Spot Size & $<$ 100 \si{\micro\meter}\\
Final RMS Bunch Length & $<$ 1 ps\\
\hline
\end{tabular}
\caption{Optimizer configuration for the DC gun UED beamline}
\label{tab:dc}
\end{table}

\begin{table}
\begin{tabular}{|y{145pt}|y{75pt}|}
\hline
{\bf Decision} & {\bf Range} \\
\hline
\hline
Bunch Charge & 0 - 300 fC\\
Initial RMS Beam Size & 0 - 50 \si{\micro\meter}\\
Intitial RMS Beam Length & 0 - 50 ps\\
MTE & 0, 150 meV\\
Gun Phase & -90 - 90 degrees\\
Peak Gun Field & 100 MV/m\\
Beam Energy & 4.5 MeV\\
Solenoid Current 1 and 2 & 0 - 4 A\\
Buncher Peak Power & 0 - 25 MW\\
Buncher Phase & 90 degrees\\
\hline
\end{tabular}

\vspace{10pt}

\begin{tabular}{|y{145pt}|y{75pt}|}
\hline
{\bf Objective} & {\bf Goal} \\
\hline\hline
RMS Emittance & Minimize\\
Delivered Bunch Charge & Maximize\\
\hline
\end{tabular}

\vspace{10pt}

\begin{tabular}{|y{145pt}|y{75pt}|}
\hline
{\bf Constraint} & {\bf Value} \\
\hline\hline
Final RMS Spot Size & $<$ 100 \si{\micro\meter}\\
Final RMS Bunch Length & $<$ 1 ps\\
\hline
\end{tabular}

\caption{Optimizer configuration for the NCRF UED beamline}
\label{tab:ncrf}
\end{table}

\begin{table}
\begin{tabular}{|y{145pt}|y{75pt}|}
\hline
{\bf Decision} & {\bf Range} \\
\hline
\hline
Bunch Charge & 100 pC\\
Initial RMS Beam Size & 0.05 - 10 mm\\
Initial RMS Beam Length & 5 - 70 ps\\
MTE & 0, 130 meV\\
Gun Gradient & 20-50 MV/m\\
Gun Phase & -60 - 60 degrees\\
Gun Energy * & 1.5-3.5 MeV\\
Solenoid 1 Field & 0 - 0.4 T\\
Capture Cavity Gradient & 0 - 32 MV/m\\
Capture Cavity Phase & -180 - 180 degrees\\
Capture Cavity Offset & 0 - 2 m\\
Solenoid 2 Field & 0 - 0.3 T\\
Solenoid 2 Offset & 0 - 2 m\\
Cryomodule Offset & 0 - 3 m\\
Accel. Cavity 1, 2, and 4 Field & 0 - 32 MV/m\\
Accel. Cavity 1, 2, and 4 Phase & -90 - 90 degrees\\
\hline
\end{tabular}

\vspace{10pt}

\begin{tabular}{|y{145pt}|y{75pt}|}
\hline
{\bf Objective} & {\bf Goal} \\
\hline\hline
RMS Emittance & Minimize\\
Final RMS Bunch Length & Minimize\\
\hline
\end{tabular}

\vspace{10pt}

\begin{tabular}{|y{145pt}|y{75pt}|}
\hline
{\bf Constraint} & {\bf Value} \\
\hline\hline
Final Energy & $>$ 90 MeV\\
Energy Spread & $<$ 200 keV\\
Higher Order Energy Spread & $<$ 5 keV\\
\hline
\end{tabular}
\caption{Optimizer configuration for the KEK gun FEL driver example.  (*) gun energy is computed from gradient and phase and not directly controlled by optimizer.}
\label{tab:fel}
\end{table}

\begin{figure*}
    \centering
    \includegraphics{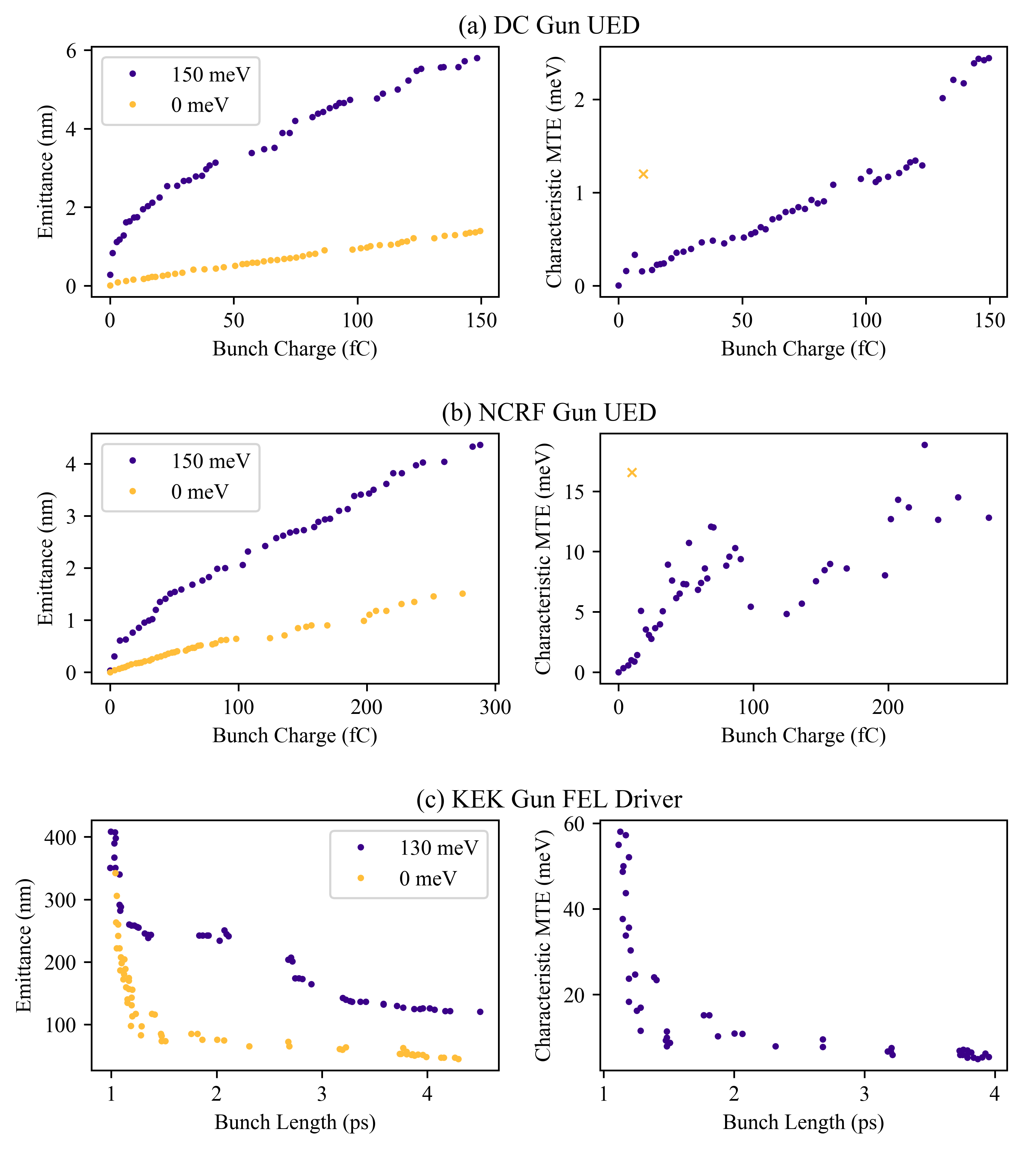}

    \caption{The Pareto fronts of each beamline for the $\sim$150 meV and 0 meV MTE photocathodes and their characteristic MTE.  The UED examples show between a factor of 10 and 100 improvement in brightness between the two Pareto fronts.  The characteristic MTE calculated from a simulation including the effects of Coulomb scattering is included for the DC and NCRF Gun UED examples as a yellow cross.}
   \label{fig:pareto_and_mte}
\end{figure*}

Initial generations of the genetic optimization were evaluated with a small number of macroparticles to develop a good approximation of the global optima before moving on to the more accurate simulations involving $10^5$ macroparticles for the UED examples and $10^4$ macroparticles for the FEL driver.
The optimization stopping condition was that improvement of the Pareto front with each successive generation fell below a threshold of approximately 10\% relative change.
The products of these optimizations are shown in Fig. \ref{fig:pareto_and_mte}.

\begin{figure}[htp]
    \centering
    \centering
    \includegraphics[width=\linewidth]{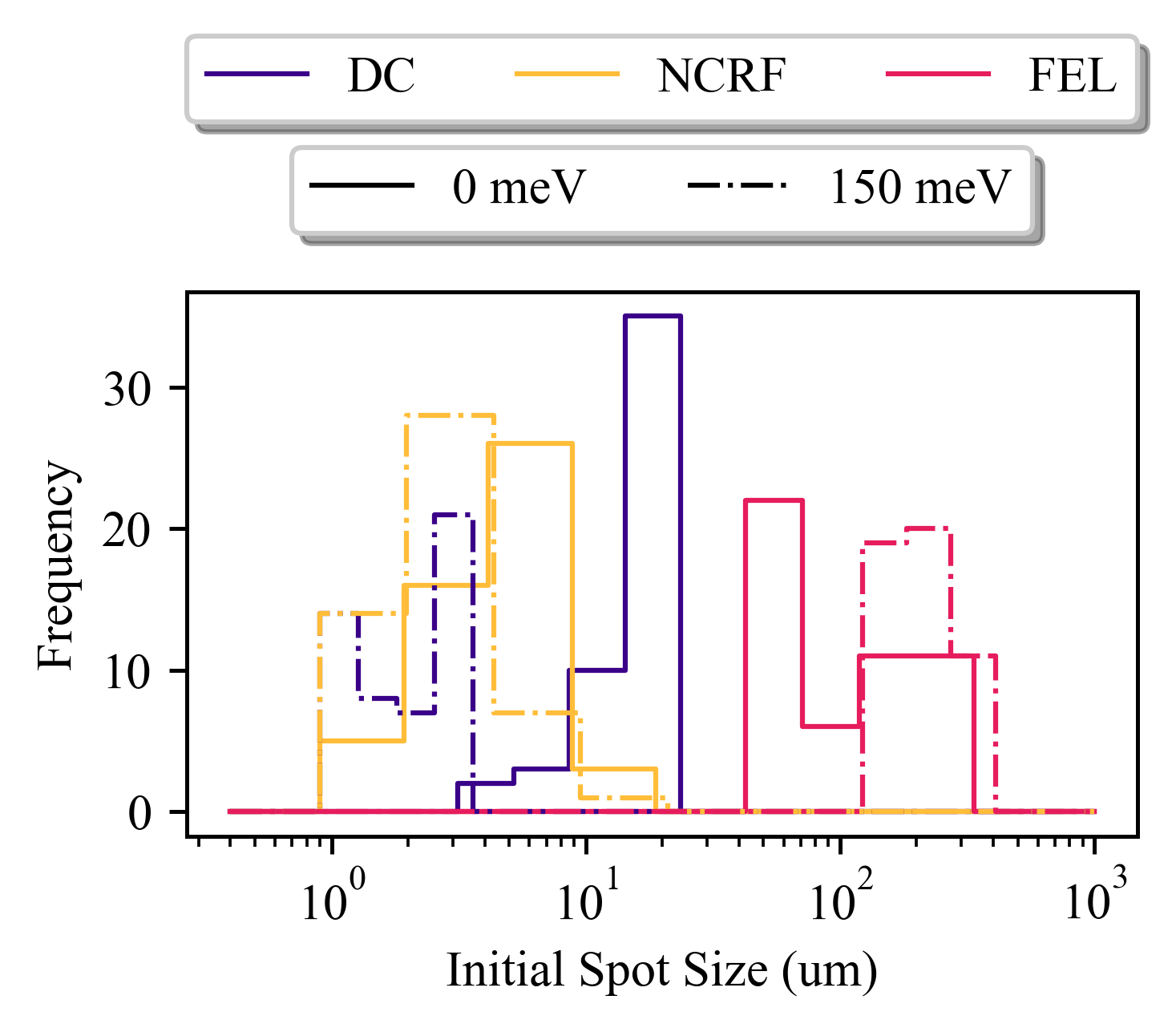}
    
    \caption{The distribution of initial spot sizes among the optimized individuals.  The three example beamlines are labeled by color and individuals from the $\sim$150 meV fronts are in dashed lines while the individuals from the 0 meV fronts are represented by solid lines.}
    \label{fig:spot_size}
\end{figure}

\begin{figure*}
    \centering
    \includegraphics{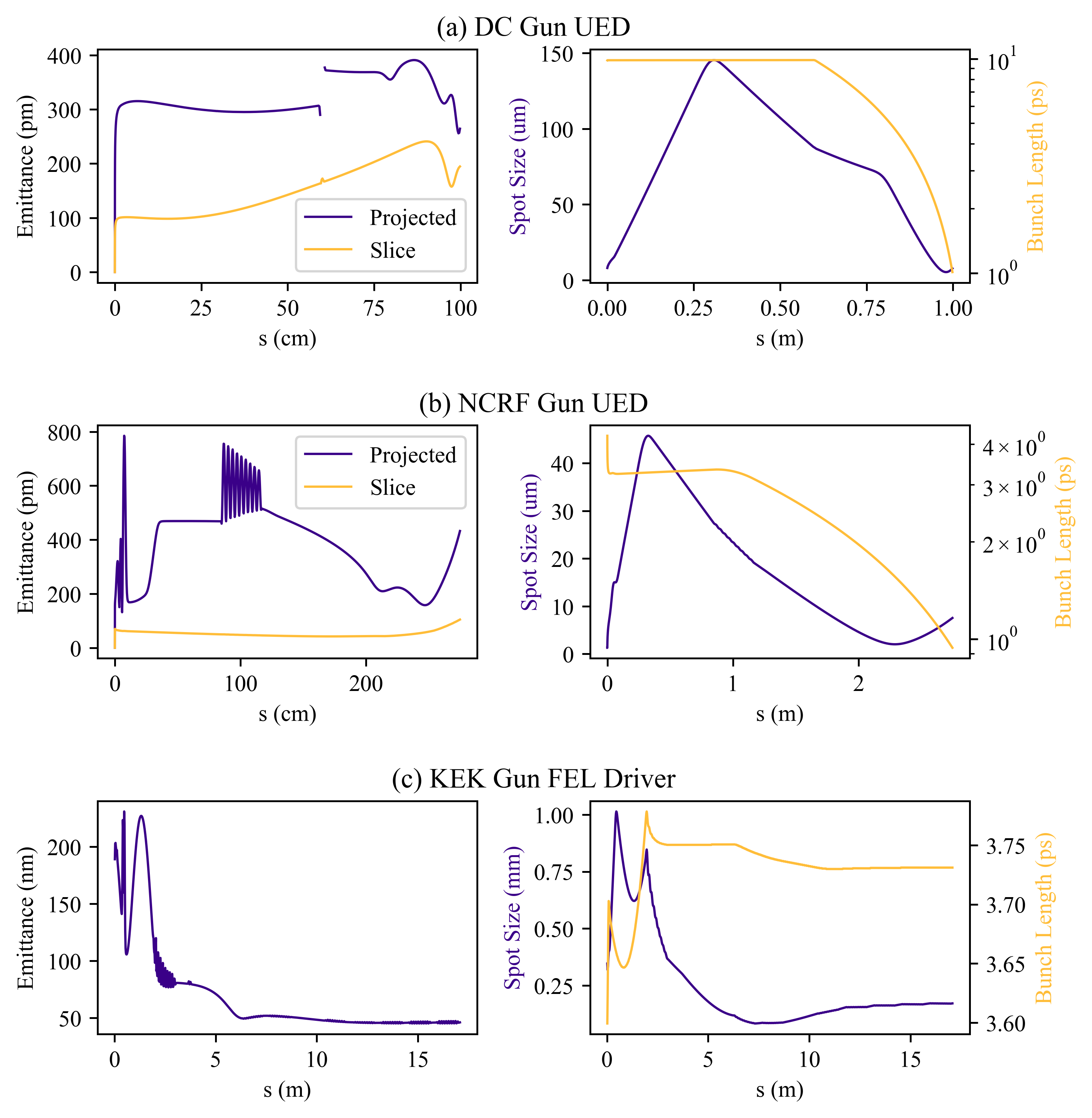}

    \caption{Emittance and beam sizes for an individual along the 0 meV Pareto front of each example. The projected emittance is the typical RMS normalized transverse emittance and the slice emittance is the average of the emittance evaluated over 100 longitudinal slices.  Beam width and length are also plotted for reference.  The total projected emittance in Fig. a is clipped at 500 pm for clarity.}
    \label{fig:example_individuals}
\end{figure*}

Both UED beamlines show a factor of between 10 and 100 improvement in brightness when the 150 meV photocathode is replaced by its 0 meV counterpart.
The degree of improvement is slightly greater in the case of the DC gun UED beamline.
As seen in Fig. \ref{fig:spot_size}, the optimizer chooses a smaller initial spot size for the NCRF gun individuals than for the DC gun individuals.
We conjecture that this is enabled by the higher accelerating gradient of the NCRF gun limiting the effects of space-charge emittance growth.
Due to the fact that initial emittance depends on both the photocathode's MTE and the initial spot size, a smaller initial spot size can mitigate the effects of a high thermal emittance photocathode.
The NCRF beamline also outperforms the DC beamline for emittance in absolute terms at similar bunch charges further suggesting a benefit with higher gradients on the cathode.
There is a sharp rise in slice emittance while the beam is still inside the gun and at low energy seen in Fig \ref{fig:example_individuals}a and \ref{fig:example_individuals}b.
This suggests that non-linear space-charge forces play a role in the residual emittance and the higher gradient and energy of the NCRF example could explain why it outperforms the DC example.
We observed that the brightness improvement from the 0 meV photocathode was limited to a factor of ten in the case of the FEL driver.
The higher bunch charge of this application is expected to increase the role of space-charge forces in transport and could be a cause of this more modest improvement.

\section{The Characteristic MTE Metric}
As long as the beam dynamics of the system do not change significantly with the introduction of a new photocathode, we can use the heuristic relationship that non-zero initial emittance will add roughly in quadrature with the emittance due to beam transport and the final emittance will be
\begin{equation}
    \varepsilon^2 \approx \varepsilon_{\text{T}}^2 + \sigma_{x,i}^2\frac{MTE}{mc^2},
\end{equation}
where $\varepsilon_{\text{T}}$ is the emittance gained in beam transport, $\sigma_{x,i}$ is the initial spot size, and $\varepsilon_{\text{C}} = \sigma_{x,i}\sqrt{\frac{MTE}{mc^2}}$ is the initial emittance due to the photocathode and initial spot size.
To understand when the photocathode's MTE is important in the final emittance, we define a \textit{characteristic MTE} that would result in the emittance contribution of the photocathode and beam transport being equal as
\begin{equation}
    \text{MTE}_{C} = mc^2 \left(\frac{\varepsilon_{\text{T}}}{\sigma_{x,i}}\right)^2.
    \label{eq:effective_mte}
\end{equation}
The characteristic MTE is a beamline specific quantity that sets the scale for when photocathodes play a significant role in determining the final emittance of a photoinjector.
Photocathode improvements down to the characteristic MTE are likely to translate into increased usable brightness.

The characteristic MTE of each example is shown in Fig. \ref{fig:pareto_and_mte}.
Photocathode improvements down to the level of single meV MTE do affect the final emittance of each photoinjector application studied here.
The characteristic MTE of both the NCRF UED and FEL driver examples increases to roughly 50 meV at high bunch charge and short bunch length respectively.
The larger characteristic MTE of the NCRF UED example is likely due to the smaller initial spot size of the individuals.
This can be seen in Fig. \ref{fig:spot_size}.
That smaller spot size will increase the characteristic MTE for the same emittance because the initial emittance is less sensitive to photocathode parameters.
Characteristic MTE at short bunch lengths in the FEL example are primarily limited by large emittance growth in beam transport.

\begin{figure}
    \centering
    \centering
    \includegraphics[width=\linewidth]{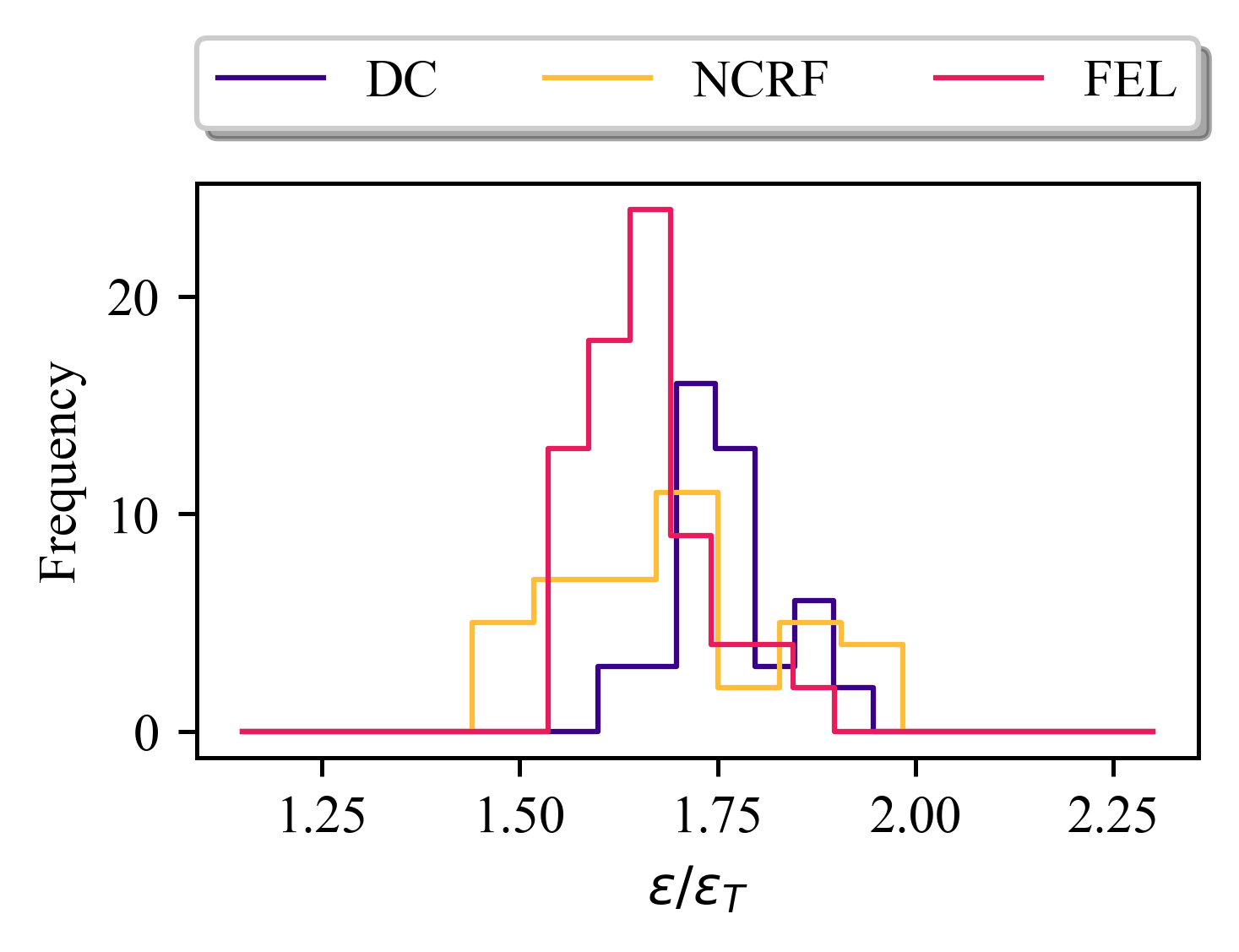}
    
    \caption{Individuals from the 0 meV beamline were re-simulated with a photocathode MTE equal to their characteristic MTE.  The frequency of the ratio of the new final emittance to the original final emittance is plotted.}
    \label{fig:ratio}
\end{figure}

To test the validity of the heuristic argument that initial and transport emittance should add in quadrature, we simulated each individual from the 0 meV Pareto fronts with a photocathode whose MTE is the characteristic MTE.
The final emittance is expected to grow by a factor of $\sqrt{2}$ and we observe the ratio to be close but slightly larger than that value.
The frequency of ratios for each beamline is plotted in Fig. \ref{fig:ratio}.
For our investigation, we assume that the insertion of a new photocathode does not significantly change beam transport.
However, this condition will be violated to some extent and could explain why the ratio observed is slightly larger than $\sqrt{2}$.

\subsection{Re-Optimization for New Photocathodes}
Our optimization experience showed that taking full advantage of the initial emittance improvements afforded by a new low MTE photocathode required the re-optimization of beamline parameters.
In particular, when individuals from the 150 meV Pareto fronts of the UED beamlines are re-simulated with a 0 meV photocathode and no changes to beamline parameters, their emittance is more than fifty percent larger than the emittance of individuals in the 0 meV Pareto front at comparable bunch charge.
This can be understood by considering the sensitivity of the transport emittance optimum to small changes in the initial spot size.

The characteristic MTE analysis does not take into account the fact that if shrinking the initial spot size from its optimal value reduces the initial emittance more than it increases emittance growth in transport, then the overall emittance will still go down.
The initial emittance, as in equation \ref{eq:initial_emittance}, can be reduced by using a smaller initial spot size.
However, if the system was already at the initial spot size which minimizes emittance growth in transport, as is the case of individuals along the 0 meV Pareto front, then changing it will negatively affect beamline performance.
Since the final emittance is roughly the quadrature sum of the initial emittance and the growth during transport, there will be a trade-off in minimizing both the initial emittance and emittance growth.
If the system was previously optimized with a high MTE photocathode, then the optimal spot size will not be at the minimum transport emittance possible and 
new low MTE photocathodes can unlock strategies the optimizer avoided due to their larger spot sizes which increase initial emittance.
In this case, re-optimization will be required upon the insertion of a new low MTE photocathode.

\begin{figure}
    \centering
    \centering
    \includegraphics[width=\linewidth]{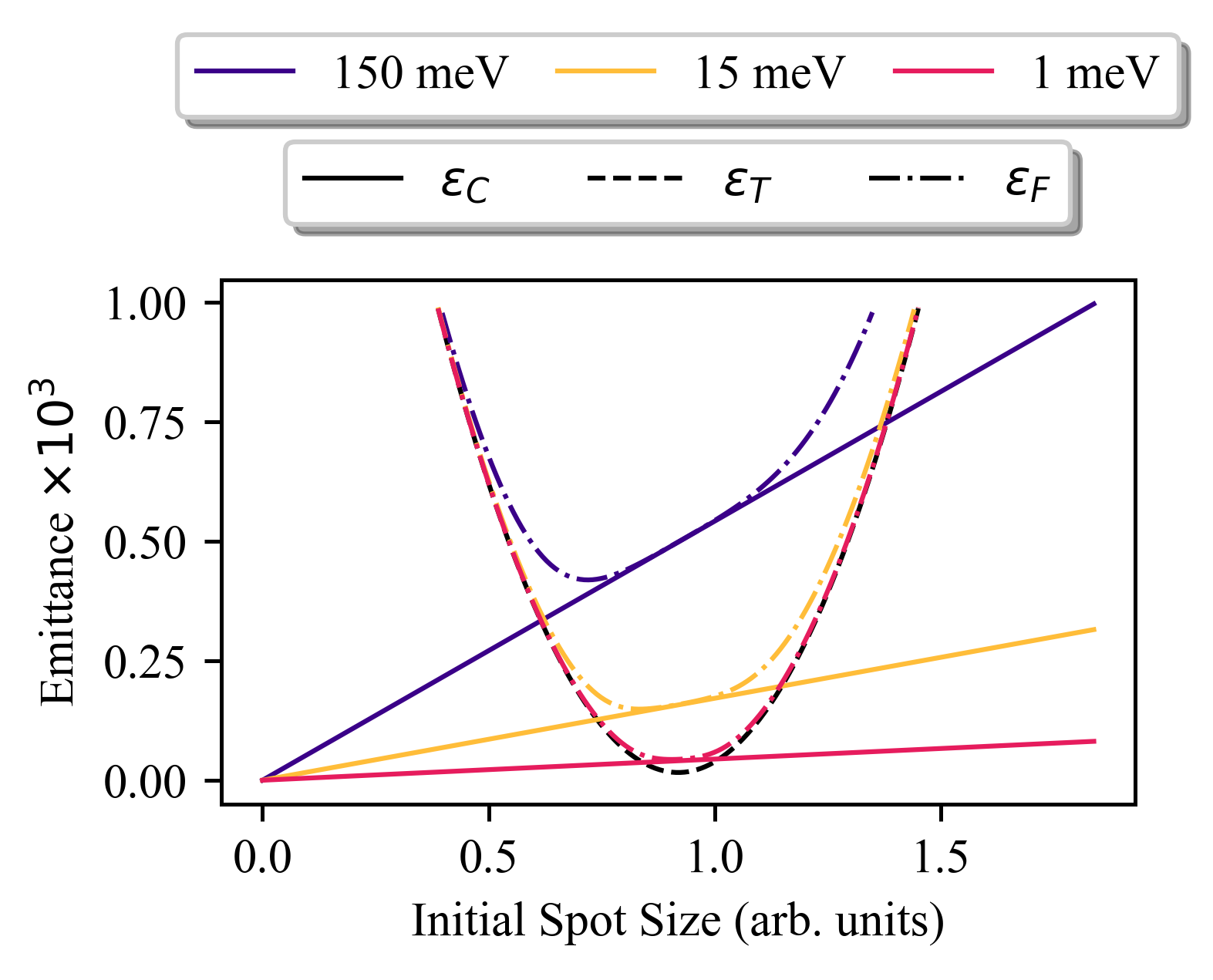}
    
    \caption{An illustration of how re-optimization may be required upon insertion of a new photocathode.  In black is the emittance due to transport ($\varepsilon_{\text{T}}$) as a function of the initial spot size.  Around the optimal spot size, $\sigma_{x,i,0}$, this is approximately quadratic. The sensitivity in this example is roughly $x\approx 0.001$.  The solid lines represent the initial emittance ($\varepsilon_{\text{C}}$) for three different thermal emittances.  The dashed lines are the final emittance ($\varepsilon_{\text{F}}$), or the quadrature sums of initial and transport emittance.  The optimal spot size with the 150 meV photocathode is significantly smaller than with a 0 meV or even 1 meV photocathode.}
    \label{fig:sensitivity}
\end{figure}

\begin{figure*}
    \centering
    \includegraphics[width=\linewidth]{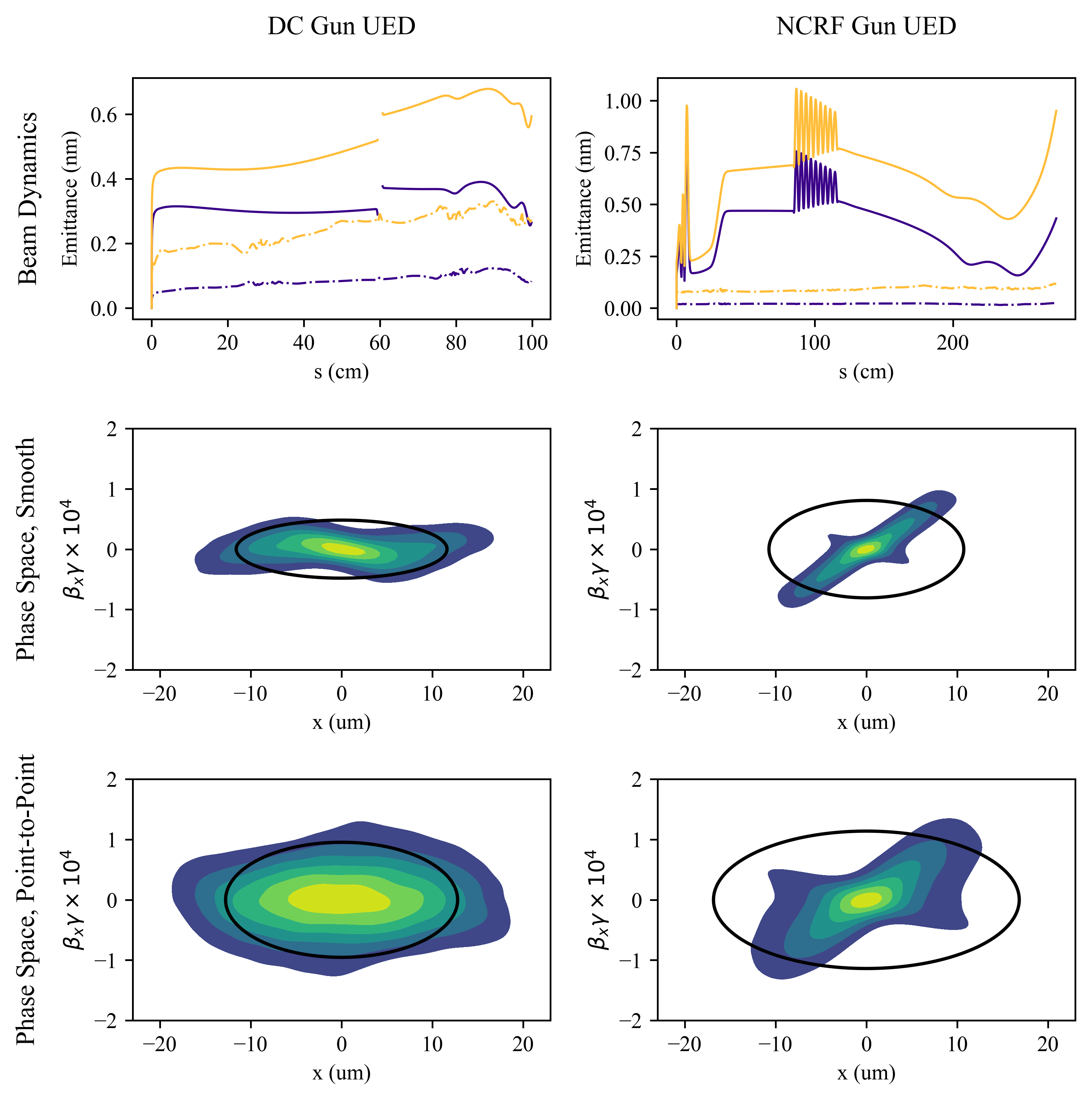}
    
    \caption{The RMS and core emittance of an individual with $10^5$ electrons per bunch from the DC gun UED and NCRF gun UED 0 meV MTE Pareto fronts.  In the row labeled "Beam Dynamics", the yellow lines were computed with the point-to-point space-charge algorithm and the blue lines with smooth space-charge.  The solid lines are the RMS normalized emittance and the dashed lines are the core emittance.  Below, are plots of the beam's transverse phase space at the sample location computed with the smooth and point-to-point methods.  Linear $x$-$p_x$ correlation have been removed and the ellipse of phase space second moments is plotted in addition to the particle density.}
    \label{fig:DIH}
\end{figure*}

This trade-off is represented graphically in Fig. \ref{fig:sensitivity} by plotting emittance as a function of initial spot size.
Initial emittance is linear in the initial spot size and is represented by a line whose slope depends on photocathode MTE.
Close to the optimum, the emittance due to transport may be expressed as a polynomial expansion in $\sigma_{x,i}$ which, to lowest order, is quadratic.
The final emittance is roughly the quadrature sum of both terms and has an optima at a smaller spot size than for transport emittance alone.
Characteristic MTE can also be represented in this plot since the initial emittance for a photocathode with an MTE equal to the characteristic MTE will pass through the vertex of the transport emittance parabola.

By using the second order expansion of beam transport's contribution to the emittance ($\varepsilon_{\text{T}}$) as a function of initial spot size around the optimum, 
\begin{equation}
    \varepsilon_{\text{T}}(\sigma_{x,i}) = A(\sigma_{x,i} - \sigma_{x,i,0})^2 + \varepsilon_{\text{T},0},
    \label{eq:optimal_emittance}
\end{equation}
we can find the new optimal emittance with non-zero MTE.
To simplify our discussion, we consider the case of optima that are highly sensitive to changes in initial spot size.
Define the unitless parameter $x = \varepsilon_{\text{T},0}/(A\sigma_{x,i,0}^2)$ to measure the optimum's sensitivity.
In the limit of sensitive optima ($x \ll 1$) the new smallest emittance when the initial spot size is allowed to vary is
\begin{equation}
\varepsilon^2_{\text{opt}} = \varepsilon_{\text{T},0}^2 + \varepsilon_{\text{C}}^2\left[1 - \frac{x}{2}\frac{\text{MTE}}{\text{MTE}_{C}}\right]\ \ \ (x \ll 1).
\label{eq:corrected_mte}
\end{equation}
The new optimal initial spot size will be smaller for the non-zero MTE photocathode and, in the limit of small $x$, is approximately
\begin{equation}
\sigma_{x,i,opt}^2 = \sigma_{x,i,0}^2\left[1 - x\frac{\text{MTE}}{\text{MTE}_{C}}\right]\ \ \ (x \ll 1).
\label{eq:corrected_sigma}
\end{equation}
In practice, we observe the tendency of the optimizer to choose smaller initial spot sizes for beamlines with non-zero photocathode MTE.
In Fig. \ref{fig:spot_size} we plot the frequency of initial spot sizes from the 0 meV and $\sim$150 meV Pareto fronts of each beamline.
For the UED examples, the initial spot sizes for individuals in the 150 meV Pareto front are universally smaller than for those in the 0 meV Pareto front.
There is less of an impact on the FEL example, which could be due to the optima being highly sensitive to changes in initial spot size.

Systems with insensitive optima (large $x$) will tolerate higher MTE photocathodes than the original characteristic MTE metric implies.
Likewise, systems where the emittance grows rapidly for small changes in $\sigma_{x,i}$ (small $x$) cannot afford to decrease the initial spot size to compensate for any increase in the photocathode MTE.
The second term in the square brackets of Eq. \ref{eq:corrected_mte} is the relative scale for how much changing the initial spot size can improve emittance and can provide a rough guide to experimentalists for determining when a new photocathode technology requires re-optimization of the beamline.
The MTE for which the transport and photocathode contributions to the final emittance are the same even when allowing the initial spot size to vary is
\begin{equation}
\text{MTE}^{\prime}_{C} = \text{MTE}_{C}\left[1 + \frac{x}{2}\right]\ \ \ (x \ll 1).
\end{equation}
Although analytical formulas for the optimal emittance and spot size which are accurate to all order in $x$ may be found, they do not lend themselves to efficient analysis and numerical methods may be better suited for investigating the properties of systems with insensitive optima.

For each system, we can use the Pareto fronts obtained for the 0 meV and $\sim$150 meV MTE photocathodes to estimate the sensitivity parameter $x$ and calculate the correction to the characteristic MTE.
These Pareto fronts give us a value of the optimal emittance from Eq. \ref{eq:corrected_mte} for two different values of $\varepsilon_{\text{C}}$ and from there we can solve for $x$.
This operation was performed on each system and the sensitivity parameter was used to calculate the corrected characteristic MTE.
The correction in all cases was at the single percent level indicating that our optima are sensitive to initial spot size.
Consequently, the \textit{uncorrected} characteristic MTE, for the three realistic photoinjectors studied here, does a good job at predicting the scale at which photocathode improvements no longer improve brightness.

\section{Stochastic Space Charge}
Disorder induced heating (DIH) is known to play a role in degrading the emittance of cold and dense electron beams.
When the distance between particles falls below the Debye length of the one component plasma, inter-particle collisions can become important and can affect the momentum distribution of the bunch in a stochastic manner.
This effect will show up prominently when the average kinetic energy of particles in the transverse direction is of the same scale as the potential energy due to the Coulomb repulsion of the particle's neighbors.
The result is that the nascent momentum spread grows above its initial value by an amount $\Delta kT [eV] = 1.04\times 10^{-9} (n_0 [m^{-3}])^{1/3}$ \cite{maxson_fundamental_2013, van_der_geer_simulated_2007}.
Using the electron number density ($n_0$) at the beginning of each optimized example, the scale of DIH expected for all three beamlines is 1 meV.
Beyond DIH, Coulomb scattering after the cathode can lead to continuous irreversible emittance growth, but these effects are difficult to estimate analytically. 
We expect DIH to be important in our simulations with 0 meV MTE photocathode due to the cold dense beams inside the guns.

To determine how much of an effect Coulomb scattering has on final emittance in our systems, one example from each of the DC and NCRF UED 0 meV Pareto fronts was chosen and simulated using a stochastic space-charge model.
The new algorithm for efficiently computing the effects of stochastic space-charge is based off of the Barnes-Hut tree method and will be discussed in detail in a forthcoming publication by M. Gordon, J. Maxson, et al.
Both the NCRF and DC UED individuals had a bunch charge of 10 fC.
Simulations were performed with GPT's smooth space-charge model discussed in \cite{bock_fast_2004} and with the tree-code method.
The RMS projected and core emittance \cite{bazarov_synchrotron_2012} along each beamline and with each space-charge model are shown in Fig. \ref{fig:DIH}.
Coulomb scattering contributes a factor of two increase in final emittance for both cases.

\section{Conclusion}
We have shown that characteristic MTE can be a useful tool in understanding the scale of MTE at which photocathode improvements translate to an increase in usable brightness.
These beamlines, which are representative of high brightness photoinjector applications, have characteristic MTEs on the scale of single to tens of meV, well below the 150 meV MTE of today's commonly used photocathodes.
Improvements in photocathode technology down to the level of 1 meV and below stands to improve the brightness of practical photoinjectors by an impressive two orders of magnitude.
However, it is not enough to simply insert a low MTE photocathode into an electron gun to achieve low final emittance.

To achieve this level of photoinjector performance, advanced optimization techniques like MOGA will need to be integrated into the design and tuning of future accelerators.
With the use of new photocathode technologies, further optimization may be required to take full advantage of low MTE.
The sensitivity of the optima to changes in initial spot size provides a guide for when it is necessary to re-optimize.
In addition, when in the regime of single meV photocathodes, existing models of smooth space-charge break down and the effects of Coulomb scattering become important in determining ultimate brightness.
Although the results of the present work are not affected by this problem because we are only concerned with order of magnitude changes in emittance, design tools for future accelerators may need to move to high performance point-to-point space-charge models to obtain good agreement with reality.

With the continued improvement of photocathode based electron sources and, in particular, the reduction of MTE in photocathode materials, bright beams will open up new possibilities for accelerator physics applications.
Notably, an increase in brightness would enable the time resolved characterization of biological macromolecules with UED \cite{sciaini_femtosecond_2011} as well as benefit X-ray FELs with a corresponding increase in total pulse energy benefiting a wide variety of x-ray scattering experiments in fields ranging from condensed matter physics, to chemistry, to biology \cite{abbamonte_new_2015}.
Work is already underway in understanding and beating the effects which limit photocathode MTE and in making existing low MTE photocathodes more practical for accelerator facility use \cite{karkare_effect_2011, jones_transverse_2018, feng_thermal_2015}.
Additionally, structured particle emitters have already been predicted to mitigate the emittance growth observed from disorder induced heating in the present simulations \cite{murphy_increasing_2015}.
If these photocathode improvements can be realized, then their results could provide as much as to two order of magnitude improvement in the final brightness of realistic modern photoinjectors.

\section{Acknowledgements}
This work was supported by the U.S. National Science Foundation under Award PHY-1549132, the Center for Bright Beams.  We thank the US-Japan Science and Technology Cooperation Program in High Energy Physics for providing additional travel funding.

\bibliographystyle{bst/apsrev4-2}
\bibliography{role_of_low_MTE}

\end{document}